\documentclass[aps,prd,onecolumn,groupedaddress,showpacs,nofootinbib,amssymb]{revtex4-2}
\usepackage[dvips]{graphicx}
\usepackage{amssymb}
\usepackage{amsmath}
\usepackage{graphicx,,color}
\usepackage{amsfonts}
\usepackage{bm}
\usepackage{cancel}
\usepackage{comment}
\usepackage{floatflt}
\usepackage{slashed}
\usepackage{appendix}
\usepackage{array}
\usepackage{tabularx}

\newcommand\be{\begin{equation}}
\newcommand\ee{\end{equation}}

\allowdisplaybreaks[4]

\begin{document}

\title{High-scale Mirror Standard Model Dark Matter, Dark Phase Transitions and Gravitational Waves Implications}
\author{V.K. Oikonomou$^{1,2}$}\email{voikonomou@gapps.auth.gr}
\affiliation{$^{1)}$Department of Physics, Aristotle University of
Thessaloniki, Thessaloniki 54124, Greece} \affiliation{
$^{2)}$Center for Theoretical Physics, Khazar University, 41
Mehseti Str., Baku, AZ-1096, Azerbaijan}


\tolerance=5000

\begin{abstract}
We consider a scenario for dark matter in the Universe, according
to which the dark matter sector is comprised by a dark Standard
Model sector which interacts only gravitationally with the
ordinary Standard Model sector. This dark Standard Model sector is
assumed to have the same symmetries as the ordinary Standard
Model, with the couplings and the scale of the mirror Standard
Model sector being different than the ordinary Standard Model
sector. Specifically, the scale of the mirror Standard Model
sector will be assumed to be quite higher compared to the ordinary
Standard Model. Also the Yukawa couplings among the mirror Higgs
and the mirror fermions are assumed to be different from those of
the Standard Model and we examine the effects of the different
scale and of the different Yukawas on the evolution of the
Universe. As we show, a mirror world phase transition occurs at
high temperatures of the baryonic Universe, which can be first
order or second order, depending on the scale of the Universe and
the Yukawa couplings. These are dark phase transitions which occur
quite earlier than the real world Standard Model electroweak phase
transition. The case of a second order phase transition is quite
interesting phenomenologically, since it can potentially have a
direct imprint on the spectrum of stochastic gravitational waves
for frequencies probed by the future gravitational wave detectors.
Also we examine whether this mirror dark matter world can form
atoms and as we show in some scenario the high scale mirror dark
matter can have both atomic and subatomic particle components. We
also give an approximation of the total equation of state of high
scale mirror DM and we discuss how high scale mirror DM can
reconcile contradicting observations like the Bullet cluster and
the Abell 520 cluster.
\end{abstract}


\maketitle

\section{Introduction}

Currently, the focus of the theoretical physics community is on
the early Universe and its mysterious physics. Many of the
upcoming experiments and observations are aiming to probe the
physics of the early Universe, via the cosmic microwave background
(CMB) \cite{CMB-S4:2016ple,SimonsObservatory:2019qwx} or also via
the stochastic gravitational wave background
\cite{Hild:2010id,Baker:2019nia,Smith:2019wny,Crowder:2005nr,Smith:2016jqs,Seto:2001qf,Kawamura:2020pcg,Bull:2018lat,LISACosmologyWorkingGroup:2022jok}.
Unfortunately, the terrestrial experiments provided nearly null
information since 2012 for fundamental physics, and the
longstanding observations of the tri-linear Higgs coupling probing
the electroweak phase transition are expected to be delivered in
about 20 years from now, so these are not promising currently. The
mysteries of the Universe currently are materialized by three
questions, whether inflation occurred, what is the nature of dark
energy and also what is the nature of dark matter (DM). The DM
problem has been thoroughly investigated by the theoretical
physics community the last 3 decades, but no definite answer has
been given regarding its nature, whether it is a particle or not.
There are alternative theories to DM theories, like the Modified
Newtonian Dynamics (MOND) theories, but these theories have no
formal relativistic description that can satisfy the GW170817
constraints, apart from the promising non-local approaches
\cite{Deffayet:2024ciu,Boran:2017rdn,Deffayet:2014lba,Deffayet:2011sk},
which are still far from providing a concrete theory that can
explain the Baryon Acoustic Observations, the CMB and other
observational features of spiral galaxies that can be explained in
an excellent way by particle dark matter. Regarding DM theories,
for 3 decades now, the focus was on Weakly Massive Interactive
Particles (WIMP), closely connected with supersymmetric extensions
of the Standard Model (SM), but no hint of the existence of a WIMP
has ever been found. And the failure to verify experimentally that
supersymmetry is an actual symmetry of subatomic physics, has
rendered the WIMP a rather vague and unrealistic candidate, along
with all the supersymmetry related theories, and string theories.
A theoretical proposal for DM which was disregarded due to the
supersymmetric WIMP searches, is the mirror DM proposal, firstly
introduced in \cite{Kobzarev:1966qya} and later developed in Refs.
\cite{Hodges:1993yb,Foot:2004pa,Berezhiani:2003wj} and also Refs.
\cite{Silagadze:2008fa,Foot:2000tp,Chacko:2005pe,Berezhiani:2000gw,Blinnikov:2009nn,Tulin:2017ara,Mohapatra:2001sx,
Blinnikov:1982eh,Blinnikov:1983gh,Foot:2016wvj,Foot:2014osa,Foot:2014uba,
Foot:2004pq,Foot:2001ft,Foot:2004dh,Foot:1999hm,Foot:2001pv,Foot:2001ne,Foot:2000iu,Pavsic:1974rq,Foot:1993yp,Ignatiev:2000yy,Ignatiev:2003js,
Ciarcelluti:2004ik,Ciarcelluti:2004ip,Ciarcelluti:2010zz,Dvali:2009fw,Foot:2013msa,Foot:2013vna,Cui:2011wk,Foot:2015mqa,Foot:2014mia,Cline:2013zca,Ibe:2019ena,
Foot:2018qpw,Howe:2021neq,Cyr-Racine:2021oal,Armstrong:2023cis,Ritter:2024sqv,Mohapatra:1996yy,Mohapatra:2000qx,Goldman:2013qla,Berezhiani:1995am,Oikonomou:2024geq,Oikonomou:2025jmy}.
The mirror DM proposal introduces the idea that the DM world is
comprised by a mirror SM world which shares the same symmetries as
our baryonic world and forms atoms, but can interact weakly with
our world or even only gravitationally. In this way, this
interacting DM world can explain several phenomenological large
scale (cluster and supercluster scales) or small scale (galactic
scale) shortcomings of cold dark matter (CDM).

In this article, we shall assume that there exists a mirror SM
world in which the Yukawa couplings of the mirror Higgs to the
mirror fermions, and also the scale of the mirror world, are
different compared to the ordinary SM world. The two sectors are
assumed to interact only gravitationally. Regarding the scale of
the mirror world, it is assumed to be higher than that of the
ordinary SM world, and also that the Yukawas are free to choose.
The mirror world has a different temperature compared to the
ordinary SM world. Furthermore, depending on the values of the
Yukawa couplings, the mirror SM world can experience a thermal
phase transition similar to the electroweak phase transition,
which occurs earlier than the electroweak phase transition and of
course at a way higher temperature. The mirror SM world
cosmological phase transition can be first or second order,
depending on the Yukawas and the scale of the mirror SM world. We
examine the phenomenological imprints of such a dark phase
transition on the spectrum of stochastic gravitational waves, and
as we show, the first order phase transition scenario cannot be
detected for a high scale mirror DM world, in contrast to a small
scale DM mirror world which can lead to detectable first order
phase transition, as it was shown in Ref.
\cite{Oikonomou:2024geq}, see also \cite{Oikonomou:2025jmy}.
However, a second order phase transition in the mirror high scale
DM world can lead to detectable imprints on the stochastic
gravitational wave spectrum, and specifically we show that such
dark second order phase transitions can be detected by the future
BBO, and DECIGO gravitational wave detectors. Another important
feature of the mirror DM world is the abundance of DM, a feature
directly related to the Yukawa couplings and the scale of the
mirror world, and depending on these critical factors, the mirror
DM can comprise all the amount of DM or some part of DM of our
Universe. We also examine in detail whether atoms can be formed in
this high scale mirror world, and also if these atoms are formed
earlier or later compared to the ordinary SM world. As we show,
atoms can indeed be formed in this mirror world, and we discuss in
the conclusions, in brief, the phenomenological implications of
this atomic mirror dark world.

\section{High Scale Mirror SM as DM and its Overall Abundance}

The high scale mirror DM we shall work with in this paper, is
essentially a copy of the SM with gauge group $G_M=SU_M(3)\times
SU_M(2)\times U_M(1)$ where the subscript ``M'' indicates the
mirror world. Thus the Universe is comprised by ordinary SM
particles with gauge group $G=SU(3)\times SU(2)\times U(1)$ and by
mirror particles with gauge group $G_M$, thus the total gauge
group of the Universe in this context is $G\otimes G_M$. It is
interesting to point that such a group structure may emerge from
the spontaneous breaking of a larger group, like for example the
$SU(4)$ which can be broken to two $SU(2)$ groups $SU(4)\to
SU(2)\times SU(2)$, but we shall not discuss in detail this
perspective, see Ref. \cite{Cembranos:2019yio} for an analysis in
another context. Thus part, or all the DM of the Universe may be
comprised by the particles belonging to the high scale mirror DM
model we assume here. These particles contribute and participate
to the Big Bang Nucleosynthesis (BBN) of the real world and affect
it, and also dark atoms can be formed. An important difference
between the SM world and the high scale mirror SM world is the
equilibrium temperature of the two worlds, which for BBN reasons
has to be different. Particularly, the mirror DM world must have a
smaller temperature compared to the SM world
\cite{Foot:2004pa,Berezhiani:2000gw}. Specifically, since the high
scale mirror DM neutrinos and mirror photons contribute to the
BBN, we would have extra neutrino species in the bounds of BBN,
that is $\delta N_{\nu}=6.14$. Let us elaborate on this and
extract the mirror world temperature, so since $\delta
N_{\nu}=6.14 \left(\frac{T'}{T}\right)^4$
\cite{Foot:2004pa,Berezhiani:2000gw}, with $T'$ and $T$ being the
mirror world and SM world equilibrium temperature, then by taking
into account the Planck constraint $\delta N_{\nu}<0.4$
\cite{Planck:2018vyg} we get $T'<0.5\, T$. Thus we shall assume
that $T'=0.5\, T$, and using this we shall calculate the total
density parameter of the mirror particles in the Universe. Note
that relativistic counting of degrees of freedom are the mirror
photons ($g=2$), the mirror neutrinos ($g=3\times 2 \times
\frac{7}{8}$) and additional relativistic mirror degrees of
freedom. The relation between the mirror SM world energy density
$\Omega_{MSM}$ and the ordinary SM world energy density
$\Omega_{SM}$ is \cite{Berezhiani:2000gw},
\begin{equation}\label{relicabundance}
\frac{\Omega_{MSM}}{\Omega_{SM}}=x^3\,D^{-K(x)}\, ,
\end{equation}
with $K(x)$ being $K(x)=\frac{1-x^2}{\sqrt{1+x^4}}$, and $D$
denotes the coefficient of the term  $\sim T^{'2}$ of the thermal
expansion of the effective potential for the high scale mirror SM
world.  It is worth further discussing how the multiplication
factor $6.14$ appears in the relation $\delta N_{\nu}=6.14
\left(\frac{T'}{T}\right)^4$.

During the BBN era, the mirror sector contains the following
relativistic degrees of freedom, first relativistic mirror
photons, secondly mirror electrons and positrons, and thirdly,
three mirror neutrino species. The total contribution of the
mirror sector to the total effective number of the neutrino
species is obtained by making a comparison between the total
mirror sector density of radiation with the energy density of a
single neutrino species, with the later being,
\begin{equation}
\label{rf21} \rho_{\nu,1} = \frac{7}{8}\frac{\pi^2}{30}\,2\,T^4
\,.
\end{equation}
Hence, the one neutrino species corresponds to,
\begin{equation}
\label{rf22} g_{\nu,1} = \frac{7}{8}\times 2 = 1.75 \,.
\end{equation}
Also, the relativistic mirror-sector degrees of freedom are the
following,
\begin{equation}
\label{rf23} g_{\gamma'} = 2 \,,
\end{equation}
\begin{equation}
\label{rf24} g_{e'^\pm} = \frac{7}{8}\times 4 = 3.5 \,,
\end{equation}
and also,
\begin{equation}
\label{rf25} g_{\nu'} = 3\times \frac{7}{8}\times 2 = 5.25 \,.
\end{equation}
Therefore, the total relativistic mirror-sector degrees of freedom
are the following,
\begin{equation}
\label{rf26} g_*' = 2+3.5+5.25 = 10.75 \, .
\end{equation}
In effect, the corresponding contribution to the total number of
effective neutrino species is,
\begin{equation}
\label{rf27} \delta N_\nu = \frac{g_*'}{1.75}
\left(\frac{T'}{T}\right)^4 \, .
\end{equation}
Hence, by substituting Eq.~(\ref{rf26}) we get,
\begin{equation}
\label{rf28} \delta N_\nu = \frac{10.75}{1.75}
\left(\frac{T'}{T}\right)^4 = 6.14 \left(\frac{T'}{T}\right)^4 \,
.
\end{equation}

As we discussed above, the mirror SM world has a smaller
temperature, hence in principle the mirror photon decoupling
occurs actually earlier to the mirror world, compared to the
ordinary SM world. This would mean that the mirror structure
formation could start earlier in the mirror world, and thus the
mirror world is being build earlier compared to the ordinary SM
baryonic world. But a crucial question is whether atoms actually
form in this mirror world, and the atom formation feature depends
on the mirror SM scale, which in our case is considered to be way
higher than the ordinary SM  electroweak scale, and on the Yukawa
couplings. At this point, let us lay out the formalism and later
on when we specify the values of the mirror SM scale and the
Yukawa couplings, we shall give detailed answers on this question,
in a model dependent way. Let us denote as usual, the ordinary SM
electroweak scale as $v=246\,$GeV, and $v_M$ the mirror SM scale.
Also let us consider the basic atom of the mirror world, the
mirror hydrogen, the binding energy of which is,
\begin{equation}\label{mirrorhydrogenbindingenergy}
{E}^{'}_B=\frac{m_{e'}{\alpha'}^2}{2}\, ,
\end{equation}
where $m_{e'}$ is the mass of the mirror electron and $\alpha'$ is
the mirror fine structure constant $\alpha'=\frac{{e'}^2}{4\pi}$,
with $e'$ being the mirror electron charge. For the ordinary SM,
the hydrogen binding energy is $E_B\sim 13.6\,$eV, and
$\alpha=1/137$, while $e=0.3$. In the mirror world, the mass of
the mirror electron is $m_{e'}=\frac{y_{e'}v_M}{\sqrt{2}}$, where
$y_{e'}$ is the mirror world Yukawa coupling of the mirror
electron to the mirror Higgs. The mirror world fine structure
constant is related to the mirror world Yukawa couplings $g'$,
$\tilde{g}'$ as follows,
\begin{equation}\label{mirrorstructurecons}
\alpha'=\frac{g'{\tilde{g}'}}{4\pi ({g'}^2+{\tilde{g}}^{'2})}\, ,
\end{equation}
hence the binding energy of the mirror hydrogen strongly depends
on the Yukawa couplings $g'$, ${\tilde{g}'}$, $y_{e'}$ and also on
the mirror scale $v_M$. For the ordinary SM, the Yukawa couplings
are quoted in Table \ref{table1} for reference and comparison with
the scenarios of the next sections.
\begin{table}[h!]
\centering
\begin{tabularx}{0.5\textwidth} {
  | >{\centering\arraybackslash}X
  | >{\centering\arraybackslash}X
   | }
 \hline
SM Yukawa \\
 \hline
$g=0.65$ \\
 \hline
 $g'=0.34\,$ \\
 \hline
$y_t=0.99$ \\
 \hline
 $y_{e'}=2.94\times 10^{-6}$ \\
 \hline
\end{tabularx}
\caption{SM Yukawa Couplings} \label{table1}
\end{table}
Hence, depending on the exact values of the mirror world scale
$v_M$ and on the Yukawa couplings, the mirror atoms can have
larger or even smaller binding energy compared to the ordinary SM
world atoms. This feature can significantly affect the
phenomenology of the high scale mirror DM model. In addition,
atoms can form earlier, or later than ordinary SM atoms can form.
In the later sections we shall address this issue concretely by
using specific values for the mirror scale and the mirror Yukawas.
But one should note that in our scenario, the binding energy of
the mirror atoms can be smaller than the ordinary world, a feature
depending on the choice of the mirror scale and the mirror
electron Yukawa coupling. Thus the nucleosynthesis and atom
formation in the mirror world can start earlier for two reasons,
first the mirror world temperature is smaller, and secondly it is
possible that for the choice of the Yukawas in the mirror world,
the binding energy for the atoms might be smaller, than in the SM
world.

Having in mind that the mirror BBN occurs actually earlier
compared to the ordinary BBN, due to the fact that the mirror
world temperature is actually smaller than the ordinary world
temperature, if atoms can form in the mirror world, then the
abundances of the mirror helium and hydrogen will actually be
different from those of the ordinary SM world. In fact, it was
shown in \cite{Berezhiani:2000gw} the ratio of the mirror helium
to mirror hydrogen will be larger, and also the baryon asymmetry
in the mirror world will be larger. We shall not analyze these
phenomenological consequences though and we will stick on the fact
that mirror DM can be particles, or particles and atoms, depending
on the scale and the Yukawas, and it is a self-interacting type of
DM.


\section{High Scale Mirror SM and the High Temperature Mirror Electroweak Phase Transition}

In this section we shall present the formalism for calculating the
finite temperature effective potential for the mirror world, which
contains all the particles of the SM, but with the scale of the
theory being significantly higher than the SM. As a working
example we shall choose the scale of the vacuum of the theory to
be of the order $\sim 7\,$TeV and specifically $v=7.786\,$TeV. As
we mentioned, this is just a working example without any
particular physical significance, regarding the numerical value of
the vacuum scale of the theory. The physical behavior of the
effective potential is similar as in the SM, but with much higher
temperatures being considered. When the temperature in the mirror
SM sector is high-we will see how much high later on this
section-the mirror $SU_M(2)$ symmetry is unbroken, and breaks
spontaneously when the temperature drops below a critical value
which depends on the choices of the Yukawas and the Higgs mass.
This mirror symmetry breaking occurs much more earlier than the
electroweak phase transition in the SM sector, because it occurs
for a way higher temperature than the $T\sim 100\,$GeV electroweak
phase transition. Let us specify the thermally corrected mirror SM
effective potential, so that we can get quantitative on the
cosmological phase transitions that might occur in the mirror SM
sector. The 1-loop effective potential of the mirror SM is equal
to \cite{Anderson:1991zb,
Quiros:1999jp,Arnold:1992rz,Carrington:1991hz,Morrissey:2012db,Dine:1992wr,Dolan:1973qd,Senaha:2020mop},
\begin{equation}\label{eq:A1}
\begin{split}
    V^{MSM}_{eff} (h', T') = & - \frac{\mu^{2}_{H'}}{2} {h'}^2 + \frac{\lambda_{H'}}{4} {h'}^4 + \sum_{i} (-1)^{F_i} n_i \frac{m^4_{i}(h')}{64 \pi^2}\left[ \ln \left( \frac{m^2_{i}(h')}{\mu^2_R}\right) - C_i \right] - \frac{n_t m^4_{t}(h')}{64 \pi^2}\left[ \ln \left( \frac{m^2_{t}(h')}{\mu^2_R}\right) - C_t \right] \\
    & + \sum_{i} \frac{n_iT'^4}{2 \pi^2} J_{B} \left(\frac{m^2_i (h')}{T'^2}\right) - \frac{n_t T'^4}{2 \pi^2} J_{F} \left(\frac{m^2_t (h')}{T'^2}\right) \\
    & + \sum_{i} \frac{\overline{n}_i T'}{12\pi} \left[m^3_i(h') - \left(M^2_i(h',T') \right)^{3/2} \right],
\end{split}
\end{equation}
where we took into account only the contributions of the mirror
Higgs, the  mirror goldstone bosons, the mirror gauge bosons and
also the contribution of the most massive quark, namely the mirror
top quark. In Eq. (\ref{eq:A1}), we denote with
 \(i = \{h', \chi', W', Z', \gamma' \}\) the
the bosons in the mirror SM sector of the theory, with the primes
indicating that we refer to mirror SM particles. The
correspondence with the SM should in principle be clear. Now
formally, by taking the high temperature expansion of the mirror
sector effective potential, by also taking into account the daisy
graphs, we get the final expression for the effective potential of
the mirror sector,
\begin{equation}\label{eq:A2}
\begin{split}
    V^{MSM}_{eff} (h', T') = & - \frac{\mu^{2}_{H'}}{2} {h'}^2 + \frac{\lambda_{H'}}{4} {h'}^4+ \frac{m^2_{h'} (h')}{24}{T'}^2 - \frac{T'}{12 \pi} \left[m^2_{h'} (h') + \Pi_{h'} (T')\right]^{3/2} + \frac{m^4_{h'}(h')}{64 \pi^2} \left[\ln \left(\frac{a_b {T'}^2}{\mu^2_R}\right) -\frac{3}{2} \right] \\
    & + \frac{3 m^2_{\chi'} (h')}{24}{T'}^2 - \frac{3T'}{12 \pi} \left[ m^2_{\chi'} (h') + \Pi_{\chi'}(T') \right]^{3/2} + \frac{ 3 m^4_{\chi'} (h')}{64 \pi^2} \left[\ln \left(\frac{a_b {T'}^2}{\mu^2_R}\right) -\frac{3}{2} \right] \\
    & + \frac{6 m^2_{W'} (h')}{24}{T'}^2 - \frac{4T'}{12 \pi} m^3_{W'} (h') - \frac{2T'}{12 \pi}\left[ m^2_{W'} (h') + \Pi_{W_L'} (T') \right]^{3/2} + \frac{ 6 m^4_{W'} (h')}{64 \pi^2} \left[\ln \left(\frac{a_b {T'}^2}{\mu^2_R}\right) -\frac{5}{6} \right] \\
    & + \frac{3 m^2_{Z'} (h')}{24}{T'}^2 - \frac{2T'}{12 \pi} m^3_{Z'} (h') - \frac{T'}{12 \pi} \left[  M^2_{Z_{L}'} (h', T') \right]^{3/2} + \frac{ 3 m^4_{Z'} (h')}{64 \pi^2} \left[\ln \left(\frac{a_b {T'}^2}{\mu^2_R}\right) -\frac{5}{6} \right] \\
    & + \frac{12 m^2_{t'} (h')}{48}{T'}^2 - \frac{ 12 m^4_{t'} (h')}{64 \pi^2} \left[\ln \left(\frac{a_f {T'}^2}{\mu^2_R}\right) -\frac{3}{2} \right] - \frac{T'}{12 \pi}\left[ M^2_{\gamma_{L}'} (h', T') \right]^{3/2},
\end{split}
\end{equation}
with $a_b=223.0993$ and $a_f=13.943$ and also the definitions of
the field-dependent masses are as follows,
\begin{equation}\label{eq:A3}
    m^2_{h'} (h') = - \mu^2_{H'} + 3\lambda_{H'} {h'}^2,
\end{equation}
\begin{equation}
    m^2_{\chi'} (h') = - \mu^2_{H'}+ \lambda_{H'} {h'}^2,
\end{equation}
and in addition,
\begin{equation}\label{effectivemassW}
    m^2_{W'} (h) = \frac{{g'}^2}{4} {h'}^2,
\end{equation}
\begin{equation}\label{effectivemassZ}
    m^2_{Z'} (h) = \frac{{g'}^2 + {\tilde{g}}^{'2}}{4} {h'}^2,
\end{equation}
\begin{equation}\label{effectivemasstop}
    m^2_{t'} (h) = \frac{y^2_{t'}}{2} {h'}^2,
\end{equation}
with \(g'\),\(\tilde{g}^{'}\) and \({y}_{t'}\) denoting the
\(SU'(2)_L\), \(U'(1)_Y\) and also mirror top quark Yukawa
couplings. Also we shall assume that the mass parameter $\mu_{H'}$
in the mirror Higgs sector is $\mu_{H'}=1208.45\,$GeV so the $\sim
{h'}^4$ mirror Higgs self-coupling $\lambda_{H'}$ has the value
$\lambda_{H'}=0.024$. For these values, the mirror Higgs mass is
$\mu_{H'}=1709\,$GeV. In addition, $T'-$dependent self-energy
corrections for the mirror Goldstone bosons and the mirror Higgs
particle are,
\begin{equation}\label{eq:A4}
    \Pi_{h'} (T') = \Pi_{\chi'} (T') = \left(\frac{3{g'}^2 }{16} + \frac{{\tilde{g}}^{'2}}{16}  +\frac{y^2_{t'} }{4} + \frac{\lambda_{H'}}{2}\right) {T'}^2
\end{equation}
and finally the mirror gauge bosons thermal masses are,
\begin{equation}\label{T-W}
    \Pi_{W_L} (T) = \frac{11}{6}{g'}^2 {T'}^2,
\end{equation}
\begin{equation}\label{Z-thermalmass}
    M^2_{Z_L} = \frac{1}{2} \left[ \frac{1}{4} \left({g'}^2 + {\tilde{g}}^{'2}\right) {h'}^2 + \frac{11}{6} \left({g'}^2 +{\tilde{g}}^{'2} \right) {T'}^2 + \sqrt{\left({g'}^2 - {\tilde{g}}^{'2} \right)^2 \left( \frac{1}{4} {h'}^2 + \frac{11}{6}  {T'}^2 \right)^2  + \frac{{g'}^2 {\tilde{g}}^{'2}}{4} {h'}^4 }\right],
\end{equation}
\begin{equation}\label{Photon-thermalmass}
    M^2_{\gamma_L} = \frac{1}{2} \left[ \frac{1}{4} \left({g'}^2 + {\tilde{g}}^{'2}\right) {h'}^2 + \frac{11}{6} \left({g'}^2 + {\tilde{g}}^{'2} \right) {T'}^2 - \sqrt{\left({g'}^2 - {\tilde{g}}^{'2} \right)^2 \left( \frac{1}{4} {h'}^2 + \frac{11}{6}  {T'}^2 \right)^2  + \frac{{g'}^2 {\tilde{g}}^{'2}}{4} {h'}^4 }\right].
\end{equation}
At this point we need to specify all the Yukawa couplings,
including the electron's Yukawa coupling which will determine the
binding energy of the mirror hydrogen atom in the high scale
mirror DM world we are discussing, and of course the mirror
electron mass itself. Regarding the values of the Yukawas, we
considered many scenarios, and depending on the values of the
Yukawas, the thermal phase transition in the mirror Higgs
effective potential can be first order or second order, and in
addition the DM abundance strongly depends on the Yukawas. We
decided to present two characteristic scenarios, which we refer to
as scenario I and scenario II.
\begin{figure}
\centering
\includegraphics[width=20pc]{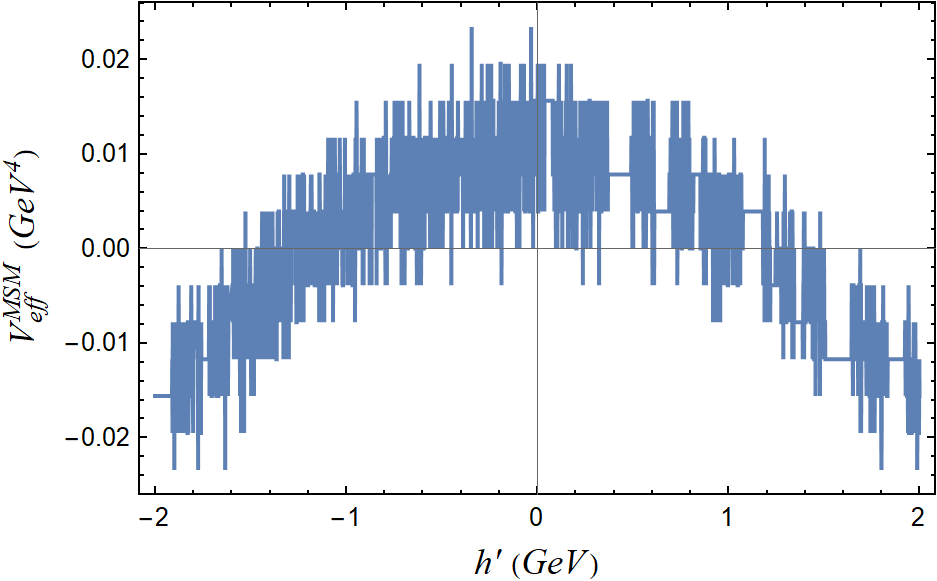}
\includegraphics[width=20pc]{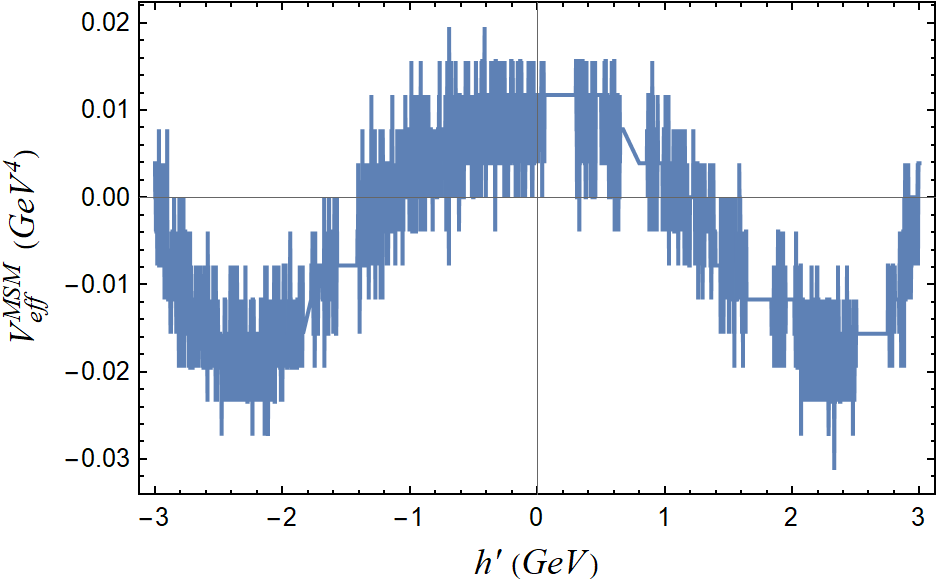}
\includegraphics[width=30pc]{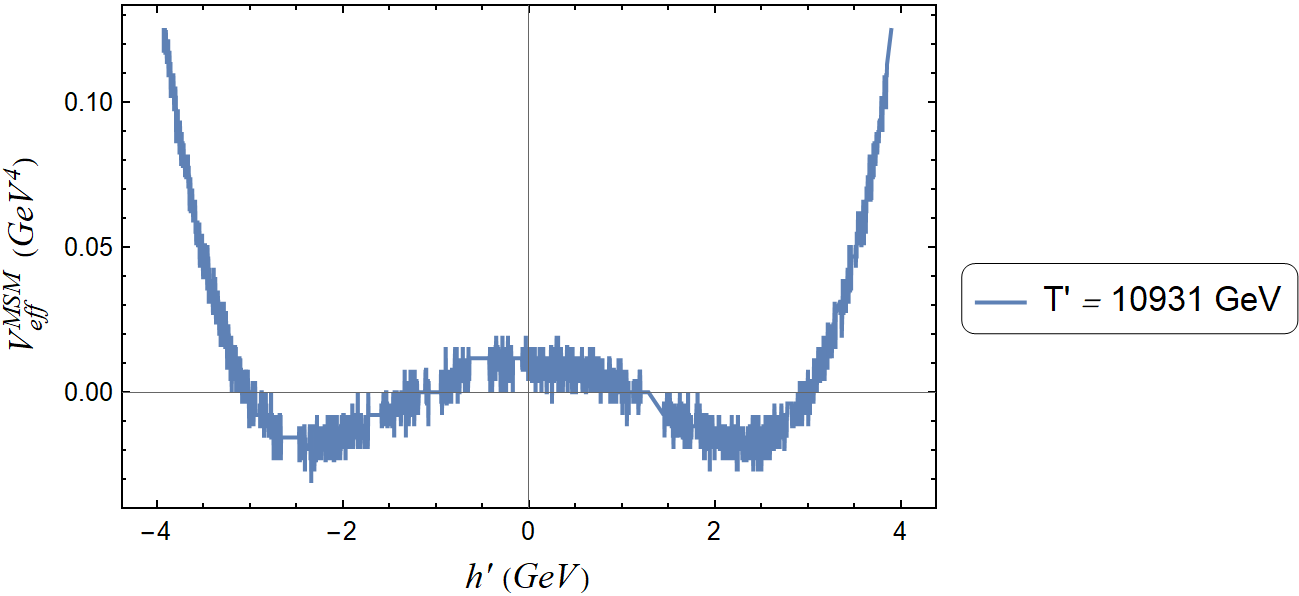}
\caption{The effective potential of the high scale mirror SM for
at exactly the critical temperature $T_c\sim 10931\,$GeV. Notice
the strong fluctuations of the effective potential at the origin
(upper plots) and the fact that the phase transition is second
order, which can be better seen in the bottom plot. Also notice in
the upper two plots the absence of a barrier, a feature that
indicates a potential second order phase transition.}\label{plot1}
\end{figure}
In the case of scenario I, the Yukawas are chosen to be smaller
than the SM case, while in scenario II the Yukawas have similar
values with the SM, but not identical. We consider the two
scenarios separately, but in both cases, the electron Yukawa
coupling is chosen to be equal to that of the SM.

\subsection{Scenario I: Second Order Phase Transition in the Mirror Electroweak Sector}

In this scenario, the Yukawas in the high scale mirror DM sector
are chosen to have smaller values compared to the SM ones, and
specifically we assume that $y_{t'}=0.0363$ $g'=0.02578$, and
\({{\tilde{g}'}}=0.02863\) and the electron Yukawa coupling is
chosen to be $y_{e'}=0.92\times 10^{-13}$. For these choices, the
masses of the mirror DM particles are quoted in Table \ref{table2}
\begin{table}[h!]
\centering
\begin{tabularx}{0.5\textwidth} {
  | >{\centering\arraybackslash}X
  | >{\centering\arraybackslash}X
   | }
 \hline
Particle & Mass (GeV) \\
 \hline
 $h'$ & $m_{h'}=1709\,$GeV \\
 \hline
 $W'$ & $m_{W'}=100.363\,$GeV \\
 \hline
 $Z'$ & $m_{Z'}=149.986\,$GeV \\
 \hline
 $t'$ & $m_{t'}=199.855\,$GeV \\
 \hline
\end{tabularx}
\caption{Mirror SM particle masses: Scenario I} \label{table2}
\end{table}
Also for these choices of the Yukawas, the fine structure constant
in the mirror world, using Eq. (\ref{mirrorstructurecons}) takes
the value $\alpha'=0.0000292$ and compare that to the ordinary SM
world value $\alpha=1/137=0.00722$. Also, for the electron Yukawa
chosen as $y_{e'}=0.92\times 10^{-13}$, the mirror electron mass
given by $m_{e'}=\frac{y_{e'}v_M}{\sqrt{2}}$ in this mirror world
is $m_{e'}\sim 0.5\,$eV, so essentially of the same order as in
the real SM world. Finally, the binding energy of the mirror
hydrogen, given by Eq. (\ref{mirrorhydrogenbindingenergy}), is
equal to ${E}^{'}_B\sim 2.16\times 10^{-10}\,$eV. So in this
mirror world, no atoms would form, even though the temperature in
the mirror sector is half of the real world temperature. This
result cannot change, unless a significantly heavy mirror electron
is used, which is a rather unwanted feature. Hence, scenario I
corresponds to a collisional DM world, which is comprised by
ordinary interacting particles, some of which are heavier than the
SM particles, like for example the Higgs. Note that previously we
mentioned that recombination in the mirror world would occur
earlier compared to the visible SM world since the mirror
temperature is smaller. However, in the case at hand this is not
true because the binding energy is extremely small, thus the
recombination in the mirror world might be significantly delayed
compared to the real world, or even will not occur at all.

\begin{figure}
\centering
\includegraphics[width=35pc]{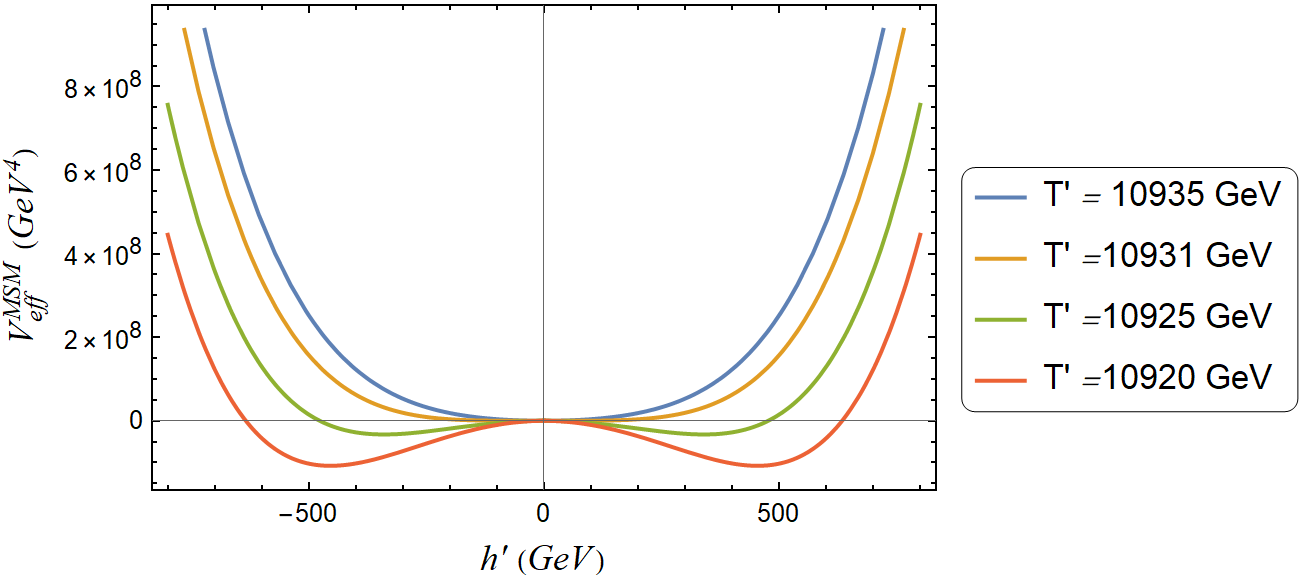}
\caption{The effective potential of the high scale mirror SM for
various temperatures near the critical temperature $T_c^{'}\sim
10931\,$GeV. The phase transition is a typical second order phase
transition.}\label{plot2}
\end{figure}

For the study of the thermal phase transition which follows, we
shall assume that the renormalization scale is $\mu_R =
2\,m_{t'}$. Let us now reveal the features of the thermal phase
transition in the high scale mirror DM world we are studying. Our
numerical analysis revealed that for the choices of the Yukawas as
in Table \ref{table2} and the choices of the scale of the theory
and the Higgs mass-or equivalently the choice of the Higgs
self-coupling $\lambda_{H'}$, the phase transition is second
order, and it starts to occur at a high critical temperature of
the order $T_c\sim 10931\,$GeV. In Figs. \ref{plot1} and
\ref{plot2} we plot the various phases of the second order phase
transition near the critical temperature and as it can be seen,
the transition is clearly second order. In Fig. \ref{plot1} we
present the behavior of the effective potential at exactly the
phase transition temperature. The important issue to note is the
strong fluctuations of the potential at the origin, typical for a
second order phase transition. In Fig. \ref{plot2} the second
order nature of the phase transition is better seen, where we plot
the effective potential for various temperatures near the critical
temperature $T_c^{'}\sim 10931\,$GeV.
The second order phase transition for the mirror Higgs scalar
field indicates that the scalar field will roll classically from
the origin towards its new minimum \cite{Athron:2023xlk} and this
can have dramatic consequences for the Universe as a whole. This
is because up to this point, that is for temperatures smaller or
equal to the critical temperature $T'\sim 10^4\,$GeV, the Universe
is deeply into the radiation domination era. In a cosmological
perturbations context, the real world temperature $T\sim 2\times
10^4\,$GeV corresponds to wavemodes of the order $k\sim
10^{10}\,$Mpc$^{-1}$, which will be probed by future gravitational
wave interferometers like LISA, DECIGO and the BBO. The fact that
at the critical point, the mirror Higgs starts rolling near its
new minimum, this will affect the background equation of state
(EoS) of the Universe, which up to that point was determined by
solely by radiation. This change in the background EoS caused by
the roll of the mirror Higgs can deform pre-existing primordial
gravitational waves for perturbation modes entering the Hubble
horizon during the epoch with temperature $T\sim 10^{4}\,$GeV.
Depending on whether the mirror Higgs rolls fast or slowly in the
origin also plays a crucial role for the determination of the
energy spectrum of the primordial gravitational waves.
\begin{figure}
\centering
\includegraphics[width=35pc]{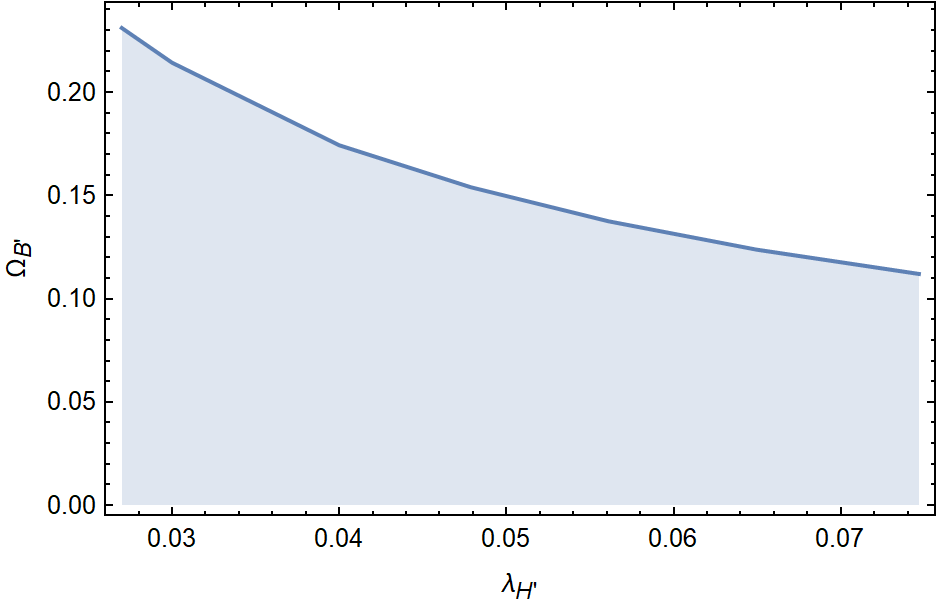}
\caption{The abundance of mirror DM particles as a function of the
mirror Higgs self-coupling $\lambda_{H'}$.}\label{plot3}
\end{figure}
This study will be presented in the next section. However all this
strongly depends on the abundance of the mirror DM, and clearly in
order for a total EoS background change at the epoch when the
temperature is of the order $T\sim 10^{4}\,$GeV, a large abundance
of mirror DM particles is needed. Let us calculate this abundance
in some detail at this point. Using the effective potential in Eq.
(\ref{eq:A2}), the field dependent term $\sim {T'}^2$ has the
following form,
\begin{equation}\label{extraeqn}
V(h',T')\sim \mathcal{D}({T'}^2-T^{'2}_0){h'}^2\, ,
\end{equation}
and in the case at hand, the parameter $D$ is equal to,
\begin{equation}\label{parameterD}
\mathcal{D}=\frac{3}{32}{g'}^2+\frac{1}{32}{\tilde{g}}^{'2}+\frac{1}{8}y_{t'}^2+\frac{\lambda_{H'}}{8}+\frac{\mu_{H'}^2}{v^2}\,
,
\end{equation}
hence, by using the numerical values for the Yukawas and the
mirror Higgs mass and self-coupling, we get, $\mathcal{D}\sim
0.0062747$. Recall that the fraction of the energy densities for
the mirror baryons $\Omega_{B'}$ over the SM baryons $\Omega_B$
is,
\begin{equation}\label{mirrorbaryonstoordinary}
\frac{\Omega_{B'}}{\Omega_B}=x^3\,\mathcal{D}^{K(x)}\, ,
\end{equation}
with $x=\frac{T'}{T}=0.64$, so with $\mathcal{D}=0.0062747$, we
get $\Omega_{B'}= 5.00491\,\Omega_{B}$. Since in the present
Universe, $\Omega_B\sim 0.05$, we finally get $\Omega_{B'}\sim
0.25$. Hence, almost all of the DM in the present Universe can be
comprised by mirror DM, and note at present day the total DM is
$\Omega_{DM}\sim 0.265$. Thus with such a large portion of DM
being composed by the mirror particles, the rolling of the mirror
Higgs will certainly affect the background EoS of the Universe.
Also, CMB constraints from Planck 2018 often favor stricter bounds
$\delta N_{\nu}<0.28$ which would yield $T'=0.3\, T$. In this case
the abundance of DM would be $\Omega_{B'}=0.13$, thus the result
of the enhancement of the gravitational waves would be weaker. But
we relied on $\delta N_{\nu}<0.4$ which also originates from the
Planck constraints. Specifically, the constraint $\Delta N_\nu
\equiv N_{\rm eff} - 3.046 \lesssim 0.4$ is not quoted explicitly
in the Planck papers, but follows directly from the reported
bounds on $N_{\rm eff}$. In \cite{Planck:2018vyg}, using
TT,TE,EE+lowE+BAO data, one finds
\begin{equation}
N_{\rm eff} = 2.99 \pm 0.17 \quad (68\% \ \mathrm{C.L.}) \, ,
\end{equation}
which implies a $95\%$ confidence interval of approximately
\begin{equation}
| \Delta N_{\rm eff} | \lesssim 2 \times 0.17 \simeq 0.34 \, .
\end{equation}
But surely, the lower value of $\Delta N_{\rm eff}$ makes the
mirror DM abundance smaller, and it is an issue which must be
reported.

For the same choice of the Yukawa couplings and the scale of the
theory, the abundance of the mirror DM particles is critically
affected by the value of the  mirror Higgs self-coupling
$\lambda_{H'}$ or equivalently the mirror Higgs mass. In Fig.
\ref{plot3} we evaluated the abundance of mirror DM particles as a
function of the mirror Higgs self-coupling $\lambda_{H'}$. As it
can be seen, as the value of the mirror Higgs self-coupling
increases, the abundance decreases. Or equivalently, as the mirror
Higgs mass increases, the abundance of mirror DM particles
decreases. However, our study revealed that the order of the phase
transition remains the same for all the values of the mirror Higgs
self-coupling used.
\begin{figure}
\centering
\includegraphics[width=35pc]{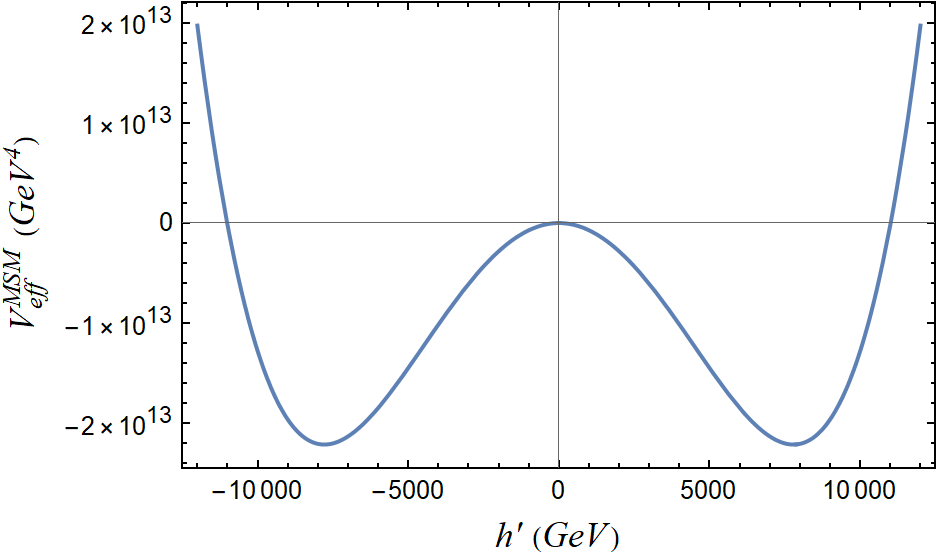}
\caption{The zero temperature effective potential of the high
scale mirror SM for scenario I.}\label{plot4}
\end{figure}
Now before closing this section, let us discuss an important
issue. The mirror sector second order phase transition occurs for
temperatures $T\sim 10^{4}\,$GeV, so quite earlier than the SM
electroweak phase transition. However, when the electroweak phase
transition occurs, the Higgs vacuum state also appears in the
Universe and thus the Universe has two, in principle, competing
vacuum states, one in each world, the mirror SM and the SM
ordinary world. Thus one could claim that the Higgs vacuum might
decay to the mirror Higgs or vice versa, depending on which is
energetically more favorable. However this is not the case in our
scenario, since the energy scale separating  the two vacua is
enormous, the mirror SM vacuum corresponds to an energy scale
$v=7886\,$GeV, while the SM Higgs vacuum $v_H=246\,$GeV, hence the
barrier that separates the two vacuum states is enormous. For the
mirror SM world, a temperature of the order $T\sim 100\,$GeV for
which the electroweak phase transition occurs, is considered as
zero temperature, thus the potential of the mirror Higgs is that
of Fig. \ref{plot4}.

The physics behind this non-transition situation between the two
vacua, due to high energy differences, is similar to the case
between the QCD vacuum which does not decay to the Higgs vacuum in
the ordinary world. Apart from this reason, in our case the mirror
world does not interact with the ordinary SM world, only
gravitationally, thus the whole transition situation cannot even
be discussed quantitatively. In a later section we shall examine
the effects of the second order phase transition we discussed in
this section on the primordial gravitational waves energy
spectrum. Before that we shall examine in the next subsection
scenario II, in which the Yukawas are chosen to take values
similar to the ones of the ordinary SM.

\subsection{Scenario II: First Order Phase Transition in the Mirror Electroweak Sector}

For the scenario II, the Yukawas are chosen to have values closer
to the ones of the SM, and specifically we assume that
$y_{t'}=0.99897$ $g'=0.62171$, and \({{\tilde{g}'}}=0.3353\) and
the electron Yukawa coupling is chosen to be the same as in the SM
$y_{e'}=2.94\times 10^{-6}$, for the same choice of the of the
mirror Higgs self-coupling $\lambda_{H'}$. For these choices, the
mirror DM particles masses are quoted in Table \ref{table3}, and
as it can be seen the mirror DM particles are quite heavier than
the SM particles.
\begin{table}[h!]
\centering
\begin{tabularx}{0.5\textwidth} {
  | >{\centering\arraybackslash}X
  | >{\centering\arraybackslash}X
   | }
 \hline
Particle & Mass (GeV) \\
 \hline
 $h'$ & $m_{h'}=1709\,$GeV \\
 \hline
 $W'$ & $m_{W'}=2420.36\,$GeV \\
 \hline
 $Z'$ & $m_{Z'}=2749.93\,$GeV \\
 \hline
 $t'$ & $m_{t'}=5499.97\,$GeV \\
 \hline
\end{tabularx}
\caption{Mirror SM particle masses: Scenario II} \label{table3}
\end{table}
Also for the choices of the Yukawas made for scenario II, the fine
structure constant in the mirror world, using Eq.
(\ref{mirrorstructurecons}) takes the value $\alpha'=0.00693$ and
compare that to the ordinary SM world value
$\alpha=1/137=0.0072229$, hence these two are similar. However,
the mirror electron Yukawa chosen as $y_{e'}=2.94\times 10^{-6}$,
the mirror electron mass which is given by
$m_{e'}=\frac{y_{e'}v_M}{\sqrt{2}}$ in this mirror world for this
scenario is $m_{e'}\sim 4.06\,$MeV, therefore quite heavier than
the real SM world electron.

Finally, the binding energy of the mirror world hydrogen, given by
Eq. (\ref{mirrorhydrogenbindingenergy}), in this case is equal to
${E}^{'}_B\sim 97.65\,$eV. Therefore, in this mirror world atoms
would form, and in fact earlier from the real world atoms. This is
for two reasons, first the binding energy is quite larger compared
to the real world, and secondly the mirror world temperature is
half of the real world temperature. Thus, this significantly
heavier mirror world we described, is heavier, it is formed much
more earlier than the ordinary world and also atoms can exist in
this mirror world. This fact, that we are basically dealing with
atomic dark matter, can have some implications for cosmological
scales, as we describe in the discussion section, later on in this
article. Now let us focus on the renormalization scale and
specifically we choose $\mu_R = 2\,m_{t'}$ and now we study the
finite temperature effective potential for the scenario II. At it
appears, the system experiences a first order phase transition in
this case, as it can also be seen in Fig. \ref{plot4} we plot the
finite temperature effective potential of the mirror SM for
various temperatures. Clearly the phase transition is a first
order one, with the critical temperature being of the order
$T^{'}_c\sim 2144\,$ at which temperature the two vacua are
equivalent. Hence, the phase transition is a supercooled one, with
the percolation temperature estimation being $T^{'}_*\sim
2000\,$GeV, at which the vacuum decay to the energetically
favorable vacuum. We need to note that this is a dark phase
transition in the mirror sector, similar in spirit but different
in context, compared to the ones appearing in the literature
\cite{Hall:2019rld,Shelton:2010ta,Dutta:2010va,Servant:2013uwa}.
The phase transition proceeds as is well described in the
literature, by bubble nucleation. This bubble collision and
percolation can lead to gravitational wave production, which will
be studied in the next section.

\begin{figure}
\centering
\includegraphics[width=35pc]{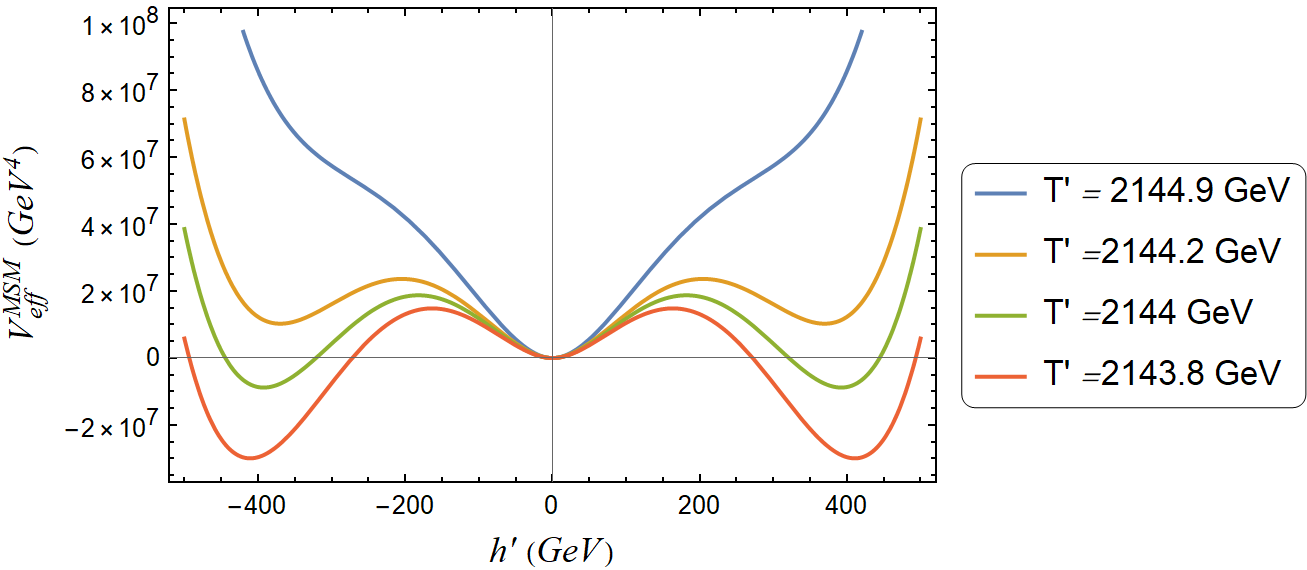}
\caption{The high temperature effective potential for the mirror
SM for the scenario II, for various temperatures near the critical
temperature. }\label{plot5}
\end{figure}

We can easily assess if the phase transition described in Fig.
\ref{plot5} is strong or not, by using the sphaleron rate
criterion,
\begin{equation}\label{sphaleron_rate}
    \frac{\upsilon_c}{T_c} > 0.6 - 1.4.
\end{equation}
In the case at hand, the fraction $\frac{\upsilon_c}{T^{'}_c}$ is
of the order $\frac{\upsilon_c}{T^{'}_c}\sim \mathcal{O}(0.186)$.
Therefore, the phase transition is deemed rather weak, but this
has to be further checked in the next section and notably it can
be even stronger if additional scalars are included in the mirror
SM sector, which can interact with the mirror Higgs.
With this weak phase transition, the departure from thermal
equilibrium is not guaranteed, so mirror world baryon asymmetry is
questionable. Hence, the mirror world also shares the SM problems
of baryon asymmetry, and perhaps singlet scalars might provide a
unique answer to this question. However this line of research is
beyond the scopes of this article. Hence, this mirror world is
formed quite earlier than the SM world, at the percolation
temperature $T'\sim 2144\,$GeV. In this case too, the mirror Higgs
vacuum occurs at a high scale, so the mirror Higgs vacuum cannot
decay to the ordinary Higgs vacuum, when the temperature of the
ordinary Universe reaches $T\sim 100\,$GeV. Also let us find the
abundance of DM in this world, in which case can also be atomic,
so from Eqs. (\ref{extraeqn}) and (\ref{parameterD}), the
parameter $D$ is $\mathcal{D}\sim 0.17051$, hence, from Eq.
(\ref{mirrorbaryonstoordinary}) we get $\Omega_{B'}\sim 0.022639$.
Hence, in this case, a very small portion of the DM in the
Universe can be high scale mirror atomic DM.

\section{Searching for Effects on Stochastic Gravitational Waves from the Dark Phase Transitions}

Having considered the possible scenarios for the dark phase
transitions that might occur in the mirror DM world, in this
section we shall investigate possible signatures of these dark
phase transitions on the primordial gravitational wave energy
spectrum. In the literature, several articles study how stochastic
gravitational waves can be generated in the early Universe
\cite{Kamionkowski:2015yta,Turner:1993vb,Boyle:2005se,Zhang:2005nw,Caprini:2018mtu,Clarke:2020bil,Smith:2005mm,Giovannini:2008tm,Liu:2015psa,Vagnozzi:2020gtf,Giovannini:2023itq,Giovannini:2022eue,Giovannini:2022vha,Giovannini:2020wrx,Giovannini:2019oii,Giovannini:2019ioo,Giovannini:2014vya,Giovannini:2009kg,Kamionkowski:1993fg,Giare:2020vss,Zhao:2006mm,Lasky:2015lej,
Cai:2021uup,Odintsov:2021kup,Lin:2021vwc,Zhang:2021vak,Visinelli:2017bny,Pritchard:2004qp,Khoze:2022nyt,Casalino:2018tcd,Oikonomou:2022xoq,Casalino:2018wnc,ElBourakadi:2022anr,Sturani:2021ucg,Vagnozzi:2022qmc,Arapoglu:2022vbf,Giare:2022wxq,Oikonomou:2021kql,Gerbino:2016sgw,Breitbach:2018ddu,Pi:2019ihn,Khlopov:2023mpo,Odintsov:2022cbm,Benetti:2021uea,Vagnozzi:2020gtf}
including the cosmological phase transitions scenarios
\cite{Apreda:2001us,Schabinger:2005ei,Kusenko:2006rh,McDonald:1993ex,Chala:2018ari,Davoudiasl:2004be,Baldes:2016rqn,Noble:2007kk,Zhou:2020ojf,
Weir:2017wfa,Hindmarsh:2020hop,Han:2020ekm,Child:2012qg,Fairbairn:2013uta,LISACosmologyWorkingGroup:2022jok,Caprini:2015zlo,Huber:2015znp,
Delaunay:2007wb,Chung:2012vg,Barenboim:2012nh,Senaha:2020mop,Grojean:2006bp,Katz:2014bha,Alves:2018jsw,Athron:2023xlk}.
We shall consider both the scenario I and scenario II cases in the
following two subsections.

\subsection{Stochastic Gravitational Waves from Scenario I}

\begin{figure}[h!]
\centering
\includegraphics[width=40pc]{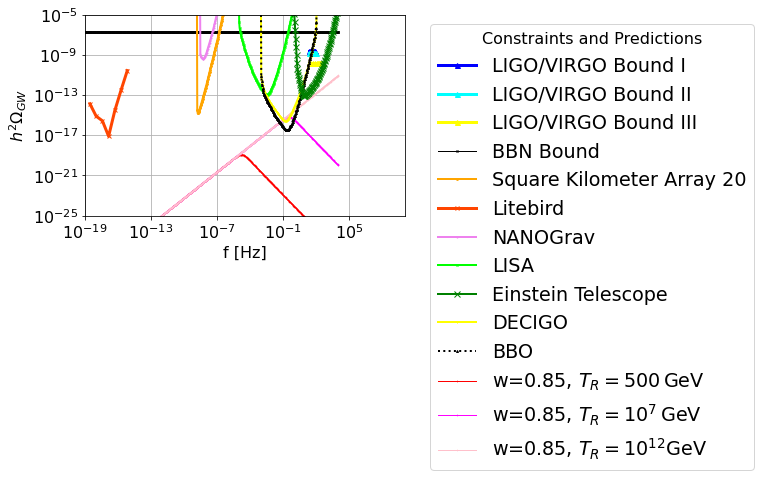}
\caption{The $h^2$-scaled gravitational wave energy spectrum for
scenario I, with the deformed background EoS having the value
$w=0.65$ corresponding to frequencies probed by LISA, BBO and
DECIGO $k_s= 10^{10}$Mpc$^{-1}$. We considered a standard
red-tilted inflationary era, with tensor spectral index being
$n_{\mathcal{T}}=-r/8$ and the tensor-to-scalar ratio being
$r=0.003$ and three distinct reheating temperatures
$T_R=500\,$GeV, $T_R=10^7\,$GeV, and $T_R=10^{12}\,$GeV.
}\label{plot7}
\end{figure}

\begin{figure}[h!]
\centering
\includegraphics[width=40pc]{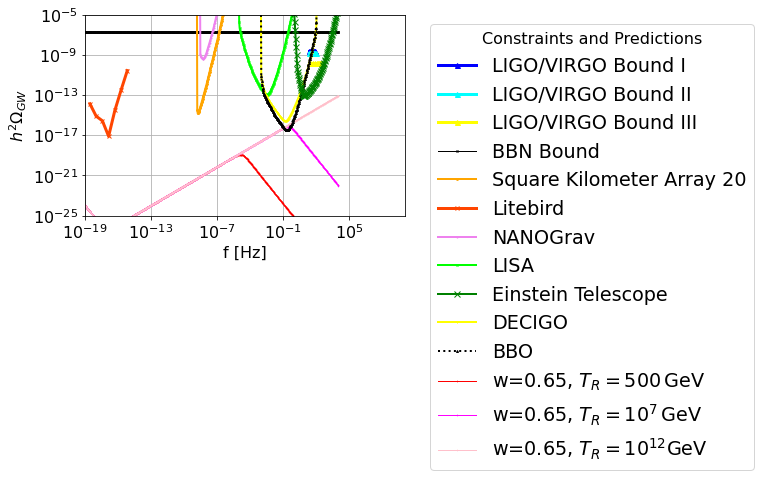}
\caption{The $h^2$-scaled gravitational wave energy spectrum for
scenario I, with the deformed background EoS having the value
$w=0.85$ corresponding to frequencies probed by LISA, BBO and
DECIGO $k_s= 10^{10}$Mpc$^{-1}$. We considered a standard
red-tilted inflationary era, with tensor spectral index being
$n_{\mathcal{T}}=-r/8$ and the tensor-to-scalar ratio being
$r=0.003$ and three distinct reheating temperatures
$T_R=500\,$GeV, $T_R=10^7\,$GeV, and $T_R=10^{12}\,$GeV.
}\label{plot8}
\end{figure}

According to the scenario I which we analyzed in the previous
section, the phase transition in the mirror DM sector is of second
order and occurs at a temperature scale $T'\sim 10^4\,$GeV, so for
a real world temperature $T\sim 2\times 10^4\,$GeV. The
cosmological modes that reenter the horizon at this era have
approximately $k\sim 10^{10}\,$Mpc$^{-1}$, therefore are probed by
LISA, DECIGO and the BBO. The second order phase transition can
occur in two ways, first the mirror Higgs can slow-roll towards
the new minimum \cite{Athron:2023xlk,Oikonomou:2023bah}, thus the
mirror Higgs EoS parameter would be somewhere in the range $-1\leq
w_{H'}\leq -1/3$, or the mirror Higgs will fast roll towards the
new minimum, thus the the mirror Higgs EoS parameter would be
somewhere in the range $1/3< w_{H'}\leq 1$\footnote{This rolling
behavior of the scalar during a second order phase transition can
also be found in the literature, see for example
\cite{Athron:2023xlk} top of page 47 in Ref.
\cite{Athron:2023xlk}.}. Recall that the second order phase
transition occurs during the radiation era, in which the total EoS
parameter is $w=1/3$. Thus this abrupt rolling of the mirror Higgs
will affect the total EoS, making it smaller or larger than the
radiation domination value $w=1/3$, depending on whether the
mirror Higgs slow-rolls or fast rolls towards the new minimum. The
reason for this radical sudden change in the total EoS is the fact
that for the scenario I, the mirror Higgs and the rest of the
mirror particles compose almost all of the DM, so a sudden change
in the mirror Higgs EoS parameter, also affects the total EoS
parameter. We shall assume that the total EoS takes the values
$w=0.25$ in the case of a slow-rolling mirror Higgs, and $w=0.65$
and also $w=0.85$ in the case of a fast rolling mirror Higgs. For
the inflationary era, we shall take that it is generated by some
standard red-tilted cosmology, like a canonical scalar field, or
some $R^2$ gravity, with the tensor spectral index being
$n_{\mathcal{T}}=-r/8$ and the tensor-to-scalar ratio being
$r=0.003$. Also we shall consider three distinct reheating
temperatures for the real world, specifically, $T_R=500\,$GeV,
hence a  scenario with low-reheating temperature, $T_R=10^7\,$GeV,
hence a scenario with intermediate-reheating temperature, and a
scenario with high reheating temperature with $T_R=10^{12}\,$GeV.
\begin{figure}[h!]
\centering
\includegraphics[width=40pc]{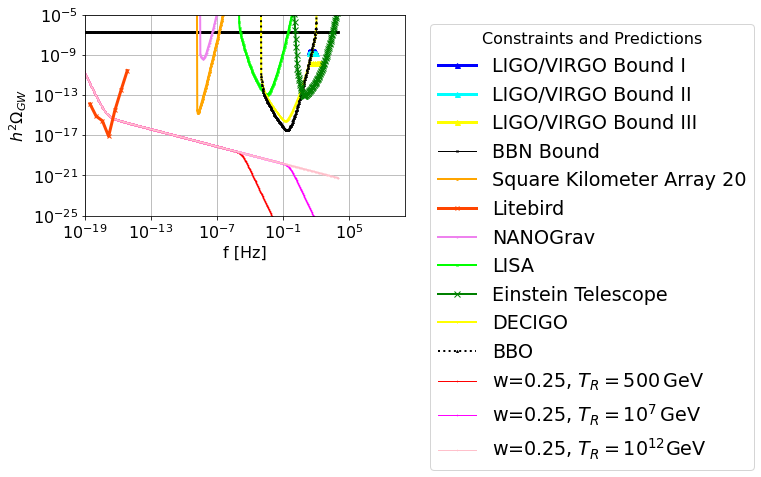}
\caption{The $h^2$-scaled gravitational wave energy spectrum for
scenario I, with the deformed background EoS having the value
$w=0.25$ corresponding to frequencies probed by LISA, BBO and
DECIGO $k_s=10^{10}$Mpc$^{-1}$. We considered a standard
red-tilted inflationary era, with tensor spectral index being
$n_{\mathcal{T}}=-r/8$ and the tensor-to-scalar ratio being
$r=0.003$ and three distinct reheating temperatures
$T_R=500\,$GeV, $T_R=10^7\,$GeV, and $T_R=10^{12}\,$GeV.
}\label{plot6}
\end{figure}
The exact effect of the total EoS deformation on the stochastic
primordial gravitational waves energy spectrum is a multiplication
factor of the form $\sim \left(\frac{k}{k_{s}}\right)^{r_c}$,
where $r_c=-2\left(\frac{1-3 w}{1+3 w}\right)$
\cite{Gouttenoire:2021jhk}, with  $k_{s}$ being the wavenumber at
which the total EoS deformation occurs, so $k_s\sim
10^{10}$Mpc$^{-1}$. Hence the $h^2$-scaled energy spectrum of the
primordial gravitational waves for the scenario I becomes,
\begin{equation}\label{GWspecfRnewaxiondecay}
\Omega_{\rm gw}(f)=S_k(f)\times
\frac{k^2}{12H_0^2}r\mathcal{P}_{\zeta}(k_{ref})\left(\frac{k}{k_{ref}}
\right)^{n_{\mathcal{T}}} \left ( \frac{\Omega_m}{\Omega_\Lambda}
\right )^2
    \left ( \frac{g_*(T_{\rm in})}{g_{*0}} \right )
    \left ( \frac{g_{*s0}}{g_{*s}(T_{\rm in})} \right )^{4/3} \nonumber  \left (\overline{ \frac{3j_1(k\tau_0)}{k\tau_0} } \right )^2
    T_1^2\left ( x_{\rm eq} \right )
    T_2^2\left ( x_R \right )\, ,
\end{equation}
where $S_k(f)$,
\begin{equation}\label{multiplicationfactor1}
S_k(f)=\left(\frac{k}{k_{s}}\right)^{r_s}\, ,
\end{equation}
with $k_{ref}$ being the CMB pivot scale
$k_{ref}=0.002$$\,$Mpc$^{-1}$, also $n_{\mathcal{T}}$ and $r$
denote the tensor spectral index of the primordial tensor
perturbations and the tensor-to-scalar ratio. In addition, $T_{\rm
in}$ stands for the horizon reentry temperature,
\begin{equation}
    T_{\rm in}\simeq 5.8\times 10^6~{\rm GeV}
    \left ( \frac{g_{*s}(T_{\rm in})}{106.75} \right )^{-1/6}
    \left ( \frac{k}{10^{14}~{\rm Mpc^{-1}}} \right )\, ,
\end{equation}
and the transfer function $T_1(x_{\rm eq})$ is equal to,
\begin{equation}
    T_1^2(x_{\rm eq})=
    \left [1+1.57x_{\rm eq} + 3.42x_{\rm eq}^2 \right ], \label{T1}
\end{equation}
with $x_{\rm eq}=k/k_{\rm eq}$ and $k_{\rm eq}\equiv a(t_{\rm
eq})H(t_{\rm eq}) = 7.1\times 10^{-2} \Omega_m h^2$ Mpc$^{-1}$,
and also the transfer function $T_2(x_R)$ is,
\begin{equation}\label{transfer2}
 T_2^2\left ( x_R \right )=\left(1-0.22x^{1.5}+0.65x^2
 \right)^{-1}\, ,
\end{equation}
where $x_R=\frac{k}{k_R}$, and also the wavenumber at the epoch
when the reheating temperature is reached, is,
\begin{equation}
    k_R\simeq 1.7\times 10^{13}~{\rm Mpc^{-1}}
    \left ( \frac{g_{*s}(T_R)}{106.75} \right )^{1/6}
    \left ( \frac{T_R}{10^6~{\rm GeV}} \right )\, ,  \label{k_R}
\end{equation}
with $T_R$ denoting the reheating temperature. Also,
$g_*(T_{\mathrm{in}}(k))$ is \cite{Kuroyanagi:2014nba},
\begin{align}\label{gstartin}
& g_*(T_{\mathrm{in}}(k))=g_{*0}\left(\frac{A+\tanh \left[-2.5
\log_{10}\left(\frac{k/2\pi}{2.5\times 10^{-12}\mathrm{Hz}}
\right) \right]}{A+1} \right) \left(\frac{B+\tanh \left[-2
\log_{10}\left(\frac{k/2\pi}{6\times 10^{-19}\mathrm{Hz}} \right)
\right]}{B+1} \right)\, ,
\end{align}
with $A$ and $B$ being equal to,
\begin{equation}\label{alphacap}
A=\frac{-1-10.75/g_{*0}}{-1+10.75g_{*0}}\, ,
\end{equation}
\begin{equation}\label{betacap}
B=\frac{-1-g_{max}/10.75}{-1+g_{max}/10.75}\, ,
\end{equation}
where $g_{max}=106.75$ and $g_{*0}=3.36$. In addition,
$g_{*0}(T_{\mathrm{in}}(k))$ can be found by using Eqs.
(\ref{gstartin}), (\ref{alphacap}) and (\ref{betacap}),with the
replacement $g_{*0}=3.36$ with $g_{*s}=3.91$. In Figs. \ref{plot6}
and \ref{plot7} and \ref{plot8} we present the $h^2$-scaled
gravitational wave energy spectrum for a deformed background EoS
cosmology with $w=0.25$ (Fig. \ref{plot6}), $w=0.65$ (Fig.
\ref{plot7}) and $w=0.85$ (Fig. \ref{plot8}) with the EoS
deformation occurring for frequencies $k_s=10^{0}$Mpc$^{-1}$. In
all the plots of Figs. \ref{plot6}, \ref{plot7} and \ref{plot8} we
used three distinct reheating temperatures $T_R=400\,$GeV,
$T_R=10^7\,$GeV, and $T_R=10^{12}\,$GeV and also a standard
red-tilted inflationary cosmology was assumed with
$n_{\mathcal{T}}=-r/8$ and $r=0.003$.
In all the plots of Figs. \ref{plot6}, \ref{plot7} and \ref{plot8}
we also included the sensitivity curves from the future
gravitational wave experiments, and in addition the constraints
coming from the BBN and the current LIGO-Virgo. As it is apparent
from Fig. \ref{plot6}, the case with $w=0.25$ which corresponds to
a slow-roll evolution in the mirror Higgs sector can be detectable
only from the Litebird experiment, while the case with $w=0.65$
can be marginally detectable from the BBO (see Fig. \ref{plot8})
while the case with $w=0.85$ can be detectable by both DECIGO and
BBO. This pattern for the primordial gravitational waves is very
unique, especially the one corresponding to the detection only by
the Literbird experiment.

\subsection{Stochastic Gravitational Waves from Scenario II}

In this section we shall consider scenario II and the possibility
of detecting a dark first order phase transition from the high
scale mirror DM theory. Let us recall how first order phase
transitions can lead to the production of stochastic gravitational
waves. We shall consider bubble collisions only, so consider a
supercooled phase transition in the context of the thin wall
approximation. During the phase transition, bubbles of the old and
the new vacuum state are ubiquitous in the Universe. The new
vacuum bubbles, have a bubble wall velocity $v_w$ and we also
assume that the new vacuum bubbles have a thin wall. These new
vacuum bubbles nucleate and grow and eventually collisions between
the old and new vacuum bubbles occur. At the point when new vacuum
bubbles grow, these are  starting to collide and eventually
coalesce, and in effect the old vacuum is transformed to the new
vacuum. The era at which bubble collision and coalescence occurs,
is exactly the era that gravitational wave production occurs.
Considering only bubble wall collisions in the envelope
approximation \cite{Kamionkowski:1993fg} and in the thin wall
limit \cite{Ellis:2020awk}, we shall make an estimation for the
energy spectrum of the produced gravitational waves for scenario
II. The energy spectrum of stochastic gravitational waves during a
first order phase transition, is given by
\cite{Ellis:2020awk,NANOGrav:2023hvm,
Xiao:2023dbb,Kamionkowski:1993fg,FitzAxen:2018vdt,Morais:2019fnm},
\begin{equation}\label{bubblecollisionenergyspectrum}
\Omega_b=\frac{\pi^2}{90}\frac{T_0^4}{M_p^2H_0^2}g_*\left(\frac{g_{*,s}^{eq}}{g_{*,s}}\right)^{4/3}\tilde{\Omega}_b\left(\frac{\alpha_*}{1+\alpha_*}\right)^2\left(
H_*R_*\right)^2\,\mathcal{S}(f/f_b)\, ,
\end{equation}
with $\tilde{\Omega}_b=0.0049$, and also the spectral function
$S(f/f_b)$ is,
\begin{equation}\label{spectralfunction}
\mathcal{S}(x)=\frac{1}{\mathcal{N}}\frac{(a+b)^c}{(b\,x^{-a/c}+a\,
x^{b/c})^c}\, ,
\end{equation}
with the positive numbers $a$ and $b$ characterizing the slope of
the spectrum at the limiting cases of low and high frequency, and
also $c$ denotes the width of the peak, with $\mathcal{N}$ being a
normalization constant,
\begin{equation}\label{mathcalN}
\mathcal{N}=\left(\frac{b}{a}
\right)^{a/n}\left(\frac{n\,c}{b}\right)^c\frac{\Gamma(a/n)\Gamma
(b/n)}{n \Gamma (c)}\, ,
\end{equation}
where $n=\frac{a+b}{c}$ and $\Gamma (z)$ stands for the gamma
function. We will use the priors of NANOGrav for $a$, $b$ and $c$,
that is, $a=1$, $b=1$, $c=3$. With regard to the rest of the
parameters in Eq. (\ref{bubblecollisionenergyspectrum}), the
parameter $\alpha_*$ is characteristic of the strength of the
phase transition. So values of the order $\alpha_*\sim
\mathcal{O}(0.01)$ characterize a weak phase transition, while of
the order $\alpha_*\sim \mathcal{O}(0.1)$ characterize an
intermediate phase transition, while of the order $\alpha_*\sim
\mathcal{O}(1)$ or even larger, characterize a strong phase
transition \cite{Athron:2023xlk}. Now, the parameters $H_*R_*$ are
related to the duration of the first order phase transition, since
$H_*R_*=(8 \pi)^{1/3}v_w\,H_*/\beta$, with $\beta$ being a
parameter characterizing directly the duration which must be
evaluated by the mirror Higgs effective potential, while $v_w$
denotes the bubble wall velocity for which we shall use NANOGrav's
value and assume $v_w=1$ \cite{NANOGrav:2023hvm}, while $H_*$
denotes the Hubble radius at exactly the percolation temperature.
Furthermore, the frequency $f_b$ that appears in Eq.
(\ref{bubblecollisionenergyspectrum}) for bubble wall collisions,
is equal to,
\begin{equation}\label{peakfrequency}
f_b\simeq 48.5\, nHz\,g_*^{1/2}\left(\frac{g_{*,s}^{e1q}}{g_{*,s}}
\right)^{1/3}\left(\frac{T_*}{\mathrm{GeV}}
\right)\frac{f_b^*R_*}{H_*R_*}\,,
\end{equation}
with $f_{b}^*=0.58/R_*$, and $R_*$ denotes the radius of the
bubble wall at exactly the percolation temperature $T_*$, and
$g_*$ denotes the number of the relativistic degrees of freedom at
the percolation temperature, $g_{*,s}$ , and also $g_{*,s}^{eq}$
denote the entropy contributing number of relativistic degrees of
freedom at the matter-radiation equality epoch. For the
calculation of the energy spectrum of stochastic gravitational
waves from first order phase transitions, the most important
parameters are $\alpha_*$ and $\beta$, and we shall calculate
these following \cite{Ellis:2020awk}. We define $\sigma$ as,
\begin{equation}\label{sigma}
\sigma=\int_{{h'}_{false}}^{h'_{true}}\mathrm{d}h'\sqrt{2V^{SM}_{eff}
(h', 0)-V_{true}}\, ,
\end{equation}
with ${h'}_{true}$ and ${h'}_{false}$ denote the mirror Higgs
particle values  at the false and true vacua respectively. Using
this, the Euclidean action $\mathcal{S}_3$ is written as,
\begin{equation}\label{euclideanaction}
\mathcal{S}_3=\frac{72\sigma^3T^{'8}}{\alpha_*^2\xi_g^4}\, ,
\end{equation}
with $\xi_g=\sqrt{\frac{30}{\pi^2g_*}}$ and also the strength of
the phase transition $\alpha_*$ can be  evaluated in the following
way,
\begin{equation}\label{strengthalpha}
\alpha_*=\frac{\Delta V}{\rho_r}\Big{|}_{T=T_*}
\end{equation}
with $\Delta V$ being the difference between the true and false
vacua in the finite temperature effective potential at the
percolation temperature. Regarding the duration of the phase
transition, it is measured by the following,
\begin{equation}\label{durationphase}
\frac{\beta}{H}=T'\frac{\mathrm{d}}{\mathrm{d}
T'}\left(\frac{S_3}{T'}
\right)=\frac{576\,\sigma^3\,{T'}^8}{\alpha_*^2\xi_g^4}\, .
\end{equation}
Let us evaluate these two parameters for the scenario II phase
transition, so at the real world percolation temperature
$T=4800\,$GeV, we have $\alpha_*=0.0000322595$ and also
$\frac{\beta}{H}\sim 10^{64}$, so the phase transition is very
weak and very slow. However, for these values of  $\alpha_*$ and
$\frac{\beta}{H}$ we found no detectable signal of gravitational
waves coming from first order phase transitions. Therefore, for
high scale mirror DM, a first order phase transition in this
sector would remain completely undetectable in the real world.
This is in contrast with the small scale mirror DM first order
phase transitions which could be detected by SKA, as it is was
shown in Ref. \cite{Oikonomou:2024geq}. We need to note that for
the calculation of the gravitational waves energy spectrum we
relied on the $S_3$ action derived in \cite{Ellis:2020awk}, which
is based on bubble collisions and the envelope approximation for
the effective potential. $S_3$ is estimated using the thin-wall
approximation, $S_3 \sim \frac{72 \sigma^3}{\alpha_*^2 T^8}$,
without solving the bounce equation explicitly. This approximation
is known to break down for weak transitions, such as the one found
in the Scenario II under study, where $\alpha_* \ll 1$. The
resulting parameter values for $\frac{\beta}{H}\sim 10^{64}$ is
clearly unphysical for the mirror world, a feature that probably
indicates a numerical instability in the temperature derivative of
the action. Regardless the signal is undetected so this feature
does not affect the phenomenology, but the consistency of the
theory must be inspected. Also, the parameter $\alpha_*$ and
$\beta$ strongly depend on the gauge, and the effective potential
is also gauge dependent. In the present context we used the Landau
gauge, which is standard in thermal field theory.  A fully
gauge-invariant treatment with using Nielsen identities or
gauge-invariant resummations, is beyond the scope of the present
work.

\section{The EoS of Mirror DM: Atomic DM and non-atomic DM and Cosmological Implications}

The importance of the high scale mirror DM model is due to the
fact that it contains both atomic and particle DM, and
specifically in the scenario II analyzed earlier. This is
important because it can reconcile distinct phenomena which
require the DM to be both cold, like the Bullet cluster, and
collisional, like the Abell 520 cluster. In this section we shall
evaluate in an approximate way the expected EoS parameter for the
high scale mirror DM which contains both atomic and particle
components, so basically only for Scenario II. The mirror sector
includes mirror photons \( \gamma' \), mirror neutrinos \( \nu'
\), and other relativistic particles like for example mirror muons
\( \mu' \), depending on the thermal history of the model. These
particles contribute to the total energy density as \(
\rho'_{\text{rel}} \sim g_i \frac{\pi^2}{30} T'^4 \), with \( g_i
\) being the number of degrees of freedom for each type of mirror
particle. Note that cold DM is pressureless, but in the case of
mirror DM, when relativistic species are considered, the pressure
for relativistic particles is \( p'_{\text{rel}} = \frac{1}{3}
\rho'_{\text{rel}} \). Now regarding the non-relativistic
components of the high scale mirror DM, for example the mirror
atoms (mirror hydrogen and so on) the energy density includes both
the rest mass and thermal contributions, as follows,
\begin{equation}\label{dmeos}
   \rho'_{\text{non-rel}} = n' m' + \frac{3}{2} n' T'
\end{equation}
where \( m' \) is the mass of the mirror atom. The pressure is \(
p'_{\text{non-rel}} \sim n' T' \) and the EoS for this component
is \( w_{\text{non-rel}} \sim \frac{T'}{m'} \ll 1 \). Now, the
total effective EoS parameter is essentially a weighted sum over
the contributions coming from the relativistic and
non-relativistic components, as follows,
\begin{equation}\label{weos}
w_{\text{eff}} = \frac{\sum_i p'_i}{\sum_i \rho'_i}
\end{equation}
where the sums run over all the mirror sector particles, atoms and
subatomic particles. For the mirror sector, if the majority of the
matter is in neutral atoms for example mirror Hydrogen, the EoS is
dominated by \( w_{\text{non-rel}} \), which in principle can be
very small for non-relativistic mirror atoms. Now, the residual
ionization fraction \( \chi(T') \) impacts the EoS significantly.
For mirror hydrogen, as the mirror temperature decreases, mirror
atoms recombine, but there is a small ionized fraction which
contributes an additional pressure. This can be approximated as
follows,
\begin{equation}\label{wesos1}
 w_{\text{eff}} \approx \chi \frac{T'}{m_{e'}} + (1 - \chi) \frac{T'}{m_{H'}}
\end{equation}
with \( \chi \) being the fraction of the ionized particles, \(
m_{e'} \) is the mirror electron mass, and \( m_{H'} \) is the
mirror hydrogen atom mass. Since \( m_{e'} \ll m_{H'} \), even a
rather small ionization fraction of the order \( \chi \sim
\mathcal{O}(10^{-3}) \) can significantly contribute to the
overall pressure, leading to a significant departure from the CDM.
Now let us quote the behavior of the mirror subatomic species, for
example the mirror muons (\( \mu' \)) and neutrinos (\( \nu' \))
with the first contributing as either relativistic or
non-relativistic particles, depending on the temperature, and the
latter contributing to the EoS with \( w = 1/3 \), which can
influence the total EoS. Regarding  mirror photons, these decouple
from the thermal bath at a temperature smaller to the ordinary SM
sector's CMB decoupling, and affect the radiation density and
ultimately the expansion rate, however their contribution to the
total EoS is relatively short-lived after the decoupling and
influences the early universe's thermal history. Also let us note
that energetic decays, for example ones involving muons and
neutrinos, like \( \mu' \to e' + \nu' + \bar{\nu}' \)) can in
principle inject energy into the mirror sector, especially during
the recombination era, and in effect could alter the thermal
history and potentially delay the recombination. These processes
can have a significant impact on both the ionization fraction and
pressure support. Now the total EoS is important because it
affects the sound speed and the Jeans scale, which are important
for cosmological observations like the Bullet cluster and Abell
520. Let us discuss in brief how the sound speed is evaluated, so
the adiabatic sound speed \( c_s \) for the mirror DM is primarily
determined by the effective EoS, as follows,
\begin{equation}\label{edsdgergd}
c_s^2 \sim w_{\text{eff}} \quad \text{(for slowly varying \( w
\))}\, ,
\end{equation}
while for mirror DM with an existing residual ionization, the
adiabatic sound speed becomes,
\begin{equation}\label{dfddfh}
   c_s \sim \sqrt{\chi \frac{T'}{m_{e'}}} \sim \sqrt{\chi} \cdot
   10^{-3}\, .
\end{equation}
The sound speed quoted above is smaller than the speed of sound in
ordinary CDM and it may impact small-scale structure formation.
Regarding the Jeans scale \( \lambda_J \), which sets the scale at
which pressure supports collapse in a self-gravitating gas, is
modified by the sound speed:
\begin{equation}\label{ergrthjgh}
   \lambda_J \sim c_s \left( \frac{\pi}{G \rho} \right)^{1/2}\, .
\end{equation}
For mirror DM, this scale could be larger compared to CDM due to
the existence of a residual pressure coming from ionized
particles, thus suppressing the formation of small scale
structures. The corresponding mass scale \( M_J \) is given by,
\begin{equation}\label{fihhuubjods}
  M_J \sim \left( \frac{c_s^2}{G} \right)^{3/2} \rho^{-1/2}\, ,
\end{equation}
so it is apparent that mirror DM behaves similarly to collisional
DM at small scales. Let us discuss now the cosmological
implications of having a combination of particle and atomic DM,
like the case of high scale DM we discussed in this paper. Due to
the pressure support coming from the residual ionization, mirror
DM may form cored profiles in galaxies, in contrast to the cuspy
profiles expected in CDM models. In addition, another feature
comes from the dissipative cooling of mirror atoms, absent in
standard CDM models, like WIMPs, since if mirror DM atoms can
radiate through mirror photons, they can cool and thus collapse,
potentially forming compact objects or disks in galaxies, and also
contributing to a diverse range of galactic structures. This is
quite novel in DM studies, usually absent in standard CDM
contexts. More importantly now, high scale DM has intriguing
collisional dynamics, and this might play an important role in
high density environments, like the Abell 520 supercluster.
Specifically, collisions or charge exchanges between mirror
particles, especially in the high-density environments, can affect
drastically the dynamics of halos and thus may allow for a degree
of self-interaction. This feature can  also lead to a more
dissipative behavior, thus in these processes, high scale mirror
DM may resemble self-interacting dark matter in dense regions.
From these arguments it is easy to understand how high scale
mirror DM can reconcile observations supporting CDM and
collisional DM, like the Bullet cluster and the Abell 520
supercluster. In the Bullet Cluster, the mirror DM behaves as
effectively collisionless, like CDM, while in environments like
Abell 520, where ionization occurs, mirror DM behaves more like
collisional DM. Specifically, in the Bullet cluster, the atomic
component of high scale mirror DM plays an important role, which
is collisionless, thus allowing the halos to pass through
unaffected. In the Abell 520 case,  mirror DM becomes partially
ionized, leading to self-interactions, causing it to lag behind
the galaxies and mimic collisional DM behavior, mimicking a gas
cloud. We need to further clarify that the difference is internal
to the mirror sector, not due to interactions with the SM. In
low-density environments the mirror atoms dominate so this
indicates a  collisionless DM behavior. In high-density
environments, partial ionization so this would indicate effective
self-interactions.

\section{Conclusions and Phenomenological Discussion}

In this article we presented a high scale mirror DM model, which
interacts only gravitationally with the SM. Depending on the
values of the Yukawa couplings, this mirror world consisting of DM
particles, can be composed by particle DM with a very large
abundance (almost all of the DM is comprised by mirror particles),
or from particle and atomic DM with a smaller abundance. We called
the former case scenario I and the latter scenario II. In both
cases, the mirror world temperature is half the ordinary world
temperature. Also in the case of scenario I, the mirror Higgs
sector undergoes a second order phase transition at a temperature
of the order $T'\sim 10^{4}\,$GeV, so twice as that in terms of
the ordinary world temperature. This mirror world phase transition
occurs deeply in the ordinary world radiation domination era, and
the second order phase transition indicates that the mirror Higgs
particle can either slow-roll or fast-roll down its potential,
towards its new vacuum state. This mirror Higgs rolling can have
imprints on the energy spectrum of the primordial gravitational
waves since the total EoS parameter will be affected by the
rolling of the mirror Higgs, especially since the mirror world
abundance is very large, almost all of the DM is comprised by
mirror particles. In the case of scenario II, a first order phase
transition occurs, which is undetectable though as we showed. Also
in the case of scenario II atoms can be formed and in fact earlier
than the ordinary world. In both cases, the dark phase transitions
occur earlier than the ordinary SM Higgs electroweak phase
transition.

It is tempting to consider the situation that part of the DM
existing in the Universe is not CDM, but it is composed by the
mirror DM we considered and the rest of the DM could be CDM, like
for example axions
\cite{Caputo:2024oqc,Kuster:2008zz,Marsh:2015xka}. Thus DM can
have both interacting and non-interacting components. The
abundance in Scenario I is strongly affected by the mirror Yukawa
couplings, so one can reduce the total abundance of mirror
particles. This could potentially have interesting
phenomenological consequences, for example CDM has several
shortcomings, such as the the diversity problem, the cusp-core
problem, and the too-big-to-fail problems. Thus having a
combination of interacting particle DM, like the high scale mirror
DM we described here, and CDM, like axions, may provide a useful
explanation for many of the CDM shortcomings, without however
altering completely the successes of the CDM description. In the
case of scenario II, atoms can be formed, thus dark compact
objects may be formed, like stars or even planets, especially
dominated by mirror helium and hydrogen and also dark galaxies,
like for example the speculated Dragonfly 44-VIRGOHI21 and
FASTJ01139+4328 which are inconsistent with the MOND theory. In
any case, it seems that the mirror DM perspective offers a
diversity of phenomenological implications that could be further
studied in the future.

\end{document}